\documentclass[12pt]{article}
\usepackage[dvips]{graphicx,color}
\def\bea{\begin{eqnarray}}
\def\eea{\end{eqnarray}}
\def\be{\begin{equation}}
\def\ee{\end{equation}}

\def\m{\mu}
\def\n{\nu}

\def\om{\omega}

\def\th{\theta}

\def\p{\partial}
\begin{document}
\title{A note on Vacuum Polarisation and Hawking Radiation}

\vspace{1cm}

\author{\sc Arundhati Dasgupta\footnote{arundhati.dasgupta@uleth.ca} and Shohreh Rahmati\footnote{shohreh.rahmati@uleth.ca}\\
Department of Physics and Astronomy, \\University of Lethbridge, Lethbridge, Canada T1K 3M4}

\maketitle

\begin{abstract}
We re-examine vacuum polarisation of a scalar field in a quasi-local volume including the horizon.  We find that
Hawking radiation rate is derived as a pure decay of vacuum due to scalar field interaction with classical gravity
exactly in the same way as the origin of vacuum polarisation effect in Electrodynamics. 
\end{abstract}

\section{Introduction}
It is often questioned whether Hawking radiation \cite{cmp} is a real physical effect or just a theoretical result due to unreal approximations.
This is prompted by the fact that for astrophysical solar mass black holes, the Hawking temperature $\sim 10^{-8} M_{\rm Sun}/M_{\rm BH}  \ ^\circ $ K
is smaller than the CMB 3$^\circ$ K. The black holes formed from stellar collapse satisfy the Chandresekhar limit and are bigger in mass than 
the sun.  These will therefore
not radiate as the ambient temperature is higher than their intrinsic Hawking temperature.
Therefore experimentally it is unlikely that one will detect this effect, unless there are primordial black holes, formed from density 
fluctuations in the early universe.
Masses of primordial black holes are small and indeed some might be evaporating away at this time. A search for primordial
black holes is yet to find results.
Surprisingly more hope comes from `analogue' systems where a similar effect can be simulated in the Lab.
A very famous example of such a system is the sonic `hole' which has apparently been tested in the Lab \cite{analogue} and more recently the
curved `Graphene sheet' \cite{graph}.
A horizon might be simulated \cite{vaz} (though there is debate whether it can be implemented in practice) using the curved 
`sheet of carbon atoms' known as graphene, and one expects Hawking radiation to emerge from the system. Given the possibility 
that Hawking radiation can be detected in the Lab, one can re-examine the physics of the phenomena and describe
it using the existing field theoretic paradigms. In the initial derivation of Hawking \cite{cmp}, a geometric approximation at the time of horizon formation
 in a collapsing situation was used to show 
a net flux of particles emergent at asymptotic future. Subsequent
derivations of particle creation in curved space-times included the Unruh effect \cite{unruh} and vacuum polarisation \cite{candelas} effects. In this note 
we will concentrate on the derivation of Hawking radiation using
a vacuum polarisation effect, which is most relevant for describing the effect lets say on a graphene sheet.

Previous computations of the vacuum polarisation of a scalar field in a black hole space-time computed the expectation value of the
scalar energy momentum tensor and usually the answers were infinite \cite{candelas}. It was then `renormalised' using a particular prescription
of subtracting the expectation value in Boulware vacuum from say the Unruh vacuum, and in the asymptotic limit it yielded the Hawking flux.
In this paper, we are interested in studying the Hawking flux near the horizon. 
We 
define the
vacuum polarisation as the expectation value of a `quasi-local action' in the vicinity of the horizon. 
Local quantities like the energy momentum tensor are infinite and one employs a point splitting method to extract finite answers \cite{dewitt}. Instead 
of using this, we use the `action' in a finite region of space and obtain the expectation value of this.
This is motivated from the Brown and York quasi-local energy concept, though we don't exactly use their prescription \cite{by}.

We show that indeed Hawking radiation flux is recovered as the vacuum expectation value of this `quasi-local action' and thus we can interpret the
Hawking effect as due to the interaction of scalar vacuaa with classical gravity which has a horizon. The vacuum relevant for this computation
is the
Unruh vacuum and the finiteness of result (not zero) can be attributed to the singularity at the horizon. 

In the next section we describe the vacuum polarisation effect for a scalar field in the vicinity of the horizon in details and section
three is a conclusion with discussions of open problems in this field.

\section{Vacuum decay in a Quasi-Local Region}
The field theory Fock space vacuum is defined as the state in which there are `no' particles. However, as was observed
by Dirac, the vacuum could also be interpreted as a sea of filled `negative energy states'. The creation 
of a particle leaves a `hole' in the Dirac sea, interpreted as the anti-particle. Thus it is easy to expect that
this `vacuum' will have a non-trivial behavior. As field theory fluctuations happens pair creation
and annihilation occurs instantaneously in vacuum, and the vacuum has `virtual particles'. This non-triviality of vacuum is 
manifest as the Casimir effect. Presence of an external field
can polarise such virtual pairs, and using the available energy, the real particles and antiparticles
can emerge causing detectable effects. This is known as vacuum polarisation. 
The most discussed  `vacuum polarisation' occurs in Electrodynamics, where
an external electric field causes the vacuum to `decay' \cite{schwing}. The exact decay rate 
for this can be computed using the vacuum expectation of the
$S$ matrix
\be
|<0|S|0>|^2 = \exp (- \int d^4 x  \ w(x)) =\exp(-W)
\label{eqn:vaccuum}
\ee
$w(x)$ is the rate of pair creation per unit time per unit volume and $W=\int d^4 x  \ w(x)$.
Obviously if $W$ is non-zero and positive it signifies a `decay'.

The S-matrix which encodes the interaction, in case of Electrodynamics, is
\be
S= \exp(\frac{\imath}{\hbar}\int d^4 x  j^{\mu}A_{\mu})
\ee
$ j^{\mu}$ is the operator scalar/fermion matter field current and $A_{\mu}$ is the vector potential corresponding to a
constant classical electric field.

By expanding the exponential in (\ref{eqn:vaccuum})
\be
W=  \frac{2 {\rm Im}}{\hbar} <0|\int d^4 x  j^{\mu} A_{\mu}|0>.
\label{eqn:image}
\ee

This was computed in \cite{schwing} and it was crucial that the vacuum expectation of the interaction term 
had a `imaginary' term for the vacuum to decay. We can easily extend the above formula to compute the
vacuum polarisation of scalar/fermions in presence of an external classical gravitational field.
As matter couples to the gravitational field through the energy momentum tensor;
the interaction is
\be
\int d^4 x \sqrt{-g}\  g^{\m \n}\  T_{\m \n}
\ee
where $T_{\m \n}$ is the scalar/fermion energy momentum tensor and $g_{\m \n}$ the metric of space-time, and g its determinant. 
The rate of vacuum decay would be
\be
W_{\rm grav}= \frac{2 {\rm Im}}{\hbar} <0|\int d^4 x \sqrt{-g} \ g^{\mu \n}\ T_{\mu \nu}|0>.
\label{eq:vac}
\ee

What makes this computation slightly difficult is the fact that there is no unique fock space vacuum in curved space. In ordinary 
quantum field theory, the vacuum is invariant as observed by a set of inertial observers. In case of a general covariant theory 
an observer is allowed to accelerate, and the QFT vacuum does not have a invariant meaning. Thus the expectation values will 
differ according to the particular vacuum in which the observation is being done. However, physical results can still be 
obtained in space times with Killing vectors which permit special observers, or when asymptotic properties can be used to 
give a invariant interpretation of the quantum amplitudes.
 
We are interested in the computation of such an amplitude in the background geometry of a black hole.
Hawking radiation was observed as a process of particle creation at the horizon, and it was believed that the
origin of this is in vacuum polarisation.
This (\ref{eq:vac}) amplitude has been calculated by several physicists before \cite{vac} in the background of a black hole metric, but we compute
a slightly modified version of (\ref{eq:vac}) motivated from the observations of \cite{adg}. In this paper
using Loop Quantum Gravity (LQG) it was shown that the horizon decreases in radius due to the time evolution
created by a quasi-local Hamiltonian at the horizon. This could be also interpreted as the vacuum expectation value of
`gravitational semiclassical fluctuations' interacting with the classical geometry of the black hole. The decrease in the
horizon radius agreed precisely with that expected from the radiation of a Hawking quanta. What was crucial in the above
derivation was that it was vacuum expectation value of: {\it A quasi-local energy at the horizon which was approximated
by taking limits to the horizon surface from both sides of the horizon (inside and outside).}\\

This suggested that we re-examine the previous derivation of scalar fields interacting with the
classical gravitational background at the horizon, but where the interaction term is not local but like the Brown and York stress energy tensor 
defined over a finite region \cite{by}.  We derive the interaction 
as the boundary term of the scalar action on the surface of a quasi-local volume which includes the horizon. We then take
the limit where the surface coincides with the horizon.
We study
the interaction at the boundary of a hollow cylinder, which
 has an outer boundary at $r_B=2GM +\delta$ and an inner boundary at $r_A=2GM-\delta$.  The
quasi-local stress-energy of \cite{by} however is not suitable in the calculation as it lacks the information
of the flux normal to the boundaries, and this is precisely what we show contributes to the
rate of particle creation. We thus compute what we name as the `quasi-local action' instead.
We simply take the usual scalar action and compute the value of the action over a volume of an annular cylinder 
which straddles the horizon, with one surface at $r_A$ outside the horizon and another surface $r_B$ inside the
horizon. The scalar action is,
\be
S= \int d^4 x \ \sqrt{-g} \ g^{\m \n} \p_\m \phi\p_\n \phi
\ee

In case there is a boundary defined using a normal $n_{\mu}$, the boundary term can be found upon simple partial integration
\be
S_{\rm quasi} = \int _{r_A}^{r_B} d^3 x \ \sqrt{-g}\  g^{\m \n} \ \phi\p_\m \phi \ n_\n = \int_{r_B} d^3 x \sqrt{-g} \ g^{\m \n} j_{\m}n_{\n} - \int_{r_A} d^3x \ \sqrt{-g} \ g^{\m \n} j_{\m} n_{\n}
\label{eqn:quasi}
\ee
where $j_{\mu}=\phi \partial_{\mu}\phi$.
Therefore the vacuum decay is simply
\be
W= 2 \  {\rm Im}<0|S_{\rm quasi}|0>
\ee

To compute the vacuum polarisation amplitude of this quasi-local action (\ref{eqn:quasi}) one has to find an appropriate vacuum. 
But the question is of course which vacuum as there is no unique vacuum in curved space-time. Currently there are three well defined vacuaa in use
for the Schwarzschild black hole \cite{candelas}

\noindent
(i) The Boulware Vacuum $|B>$\\ defined by requiring normal modes to be positive frequency with respect to the Killing vector $\frac{\partial}{\partial t}$ with respect to which the exterior region is static.\\
(ii) The Hartle-Hawking Vacuum $|H>$\\ defined by taking incoming modes to be positive frequency with respect to the Kruskal coordinate $V$, the canonical affine parameter on the future horizon, and outgoing modes to be positive frequency with respect to $U$ the affine parameter on the past horizon.\\
(iii) The Unruh Vacuum $|U>$\\ defined by taking modes that are incoming from $\cal I^{-}$ to be positive frequency with respect to $\frac{\partial}{\partial t}$, while those that emanate from the past horizon are taken to be positive frequency with respect to $U$, the canonical affine parameter on the past horizon.
(See Figure (\ref{fig:kruskal})).\\
Previous computations of vacuum polarisation using the gravitational energy momentum tensor uses a renormalisation prescription or a subtraction scheme to compute the
Hawking flux, e.g.
\be
<T_{\m \n}>_{\rm ren}= {\rm Lim}_{r\rightarrow \infty} <U|T_{\m \n}|U>-<B|T_{\m \n}|B>
\ee

is the renormalised Unruh vacuum expectation value of the energy momentum tensor at infinity.

The `interaction term' or the scalar coupling term is derived from the scalar action. We thus concentrate on the action
and in our computation we simply take the expectation value of the `quasi-local action'
in an appropriate vacuum. For the purposes of the calculation it is the Unruh vacuum which is most appropriate. In a collapsing situation,
particles with positive frequency wrt the generators of past horizon emerge from behind the horizon \cite{unruh}. It is these modes which
contribute to the inner surface of the cylinder.

Note we place this cylinder in a collapsing situation, in the region where the future horizon has been formed,
or the `steady region' . As the apparent horizons and the event horizons coincide in these Cauchy slices, one could take a `quasilocal volume' by taking a spherical region of the apparent horizon in one time slice $\tau$
and generate a volume extending to the next time slice $\tau + \delta \tau$. 

\begin{figure}
\begin{center}
\includegraphics[scale=0.4]{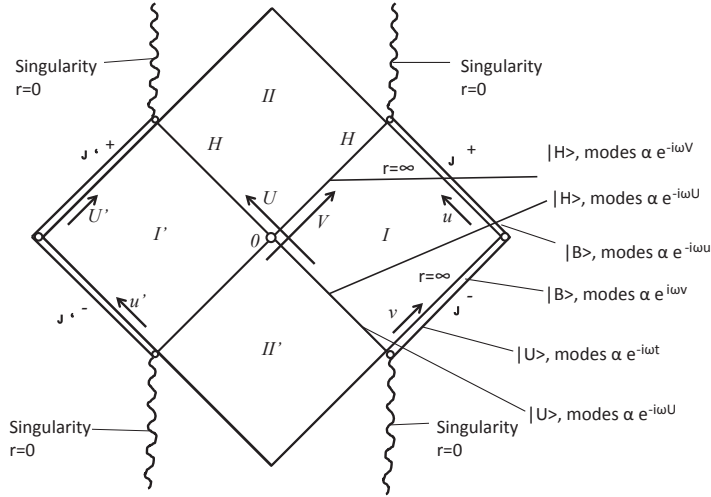}
\end{center}
\caption{The different vacua}
\label{fig:kruskal}
\end{figure}
Thus
\be
W= 2 {\rm Im}[<U| \int_{r_B} \sqrt{-g} g^{\m \n} j_{\m}n_{\n}|U>-<U|\int_{r_A} \sqrt{-g} g^{\m \n} j_{\m} n_{\n}|U>]
\label{eqn:ima}
\ee
 We take the $r_B$ as a point outside the horizon and $r_A$ as a point inside the horizon.

At a point outside the horizon $r_B$, the scalar field operator is a linear combination of the modes defined outside the horizon \cite{unruh}

To obtain the scalar field in the exterior region, the metric is defined in Schwarszchild coordinates
\be
ds^2= - \left(1-\frac{2GM}{r}\right) dt^2 + \left(1-\frac{2GM}{r}\right)^{-1} dr^2 + r^2 d\Omega
\ee
where $d\Omega= d\theta^2 + \sin^2\theta d\phi^2$. To solve for the scalar normal modes, we define
$r^*= r+ 2GM\ln\left(\frac r{2GM}-1\right)$, the metric reduces to
\be
ds^2 = -\left(1-\frac{2GM}{r}\right)[dt^2 - dr^{* 2}] + r^2 d\Omega
\ee
The scalar field equations in this background are
\bea
\frac{1}{\sqrt{-g}}\  \partial_{\mu}\  \left(\sqrt{-g}g^{\m \n}\partial_{\n}\right)\phi & =& 0\\
\partial_t^2\phi - \partial_{r^*}^2 \phi + \left(1- \frac{2GM}{r}\right)L_{\th,\phi}\phi & =& 0 
\eea
($L_{\th,\phi}$ is the angular differential operator whose eigenfunctions are the spherical harmonics $Y_{lm}(\th,\phi)$)
In the Limit $r\rightarrow 2GM$, the wavefunctions which are positive frequency wrt $t$ are well defined as there is a Killing vector $\partial_t$ 
in the exterior region. These are easily identified as $\propto e^{-i\omega t} e^{\pm i\omega r^* }$. Using the conventional notation of \cite{dewitt} we identify these as $u_{\omega lm}$ labelled using the frequency $\omega$ and angular momentum quantum numbers $l,m$: 
However unlike the `boundary conditions' used in \cite{dewitt}, we redefine them for this particular physical situation.

\be
\phi (r_B)= \sum_{\omega lm} [a_{r_B}  u_{\omega l m} + a_{r_B}^{\dagger} u^*_{\omega l m}]
\label{eqn:rb}
\ee

\be
u_{\omega l m} = \frac{1}{(4 \pi \omega)^{1/2} \ r} e^{-\iota \omega t} e^{\iota \omega r^*} \ Y_{lm} 
\label{eqn:rb1}
\ee
This particle was a `outgoing flux' at the surface of the cylinder $r_B$.

For the boundary at $r_A$, we take the vacuum corresponding to modes of the scalar field which
are positive frequency wrt the null generators of the past horizon, as these are the modes
which emerge from behind the horizon even in a collapsing case \cite{unruh}.
As the space-time behind the horizon is accessible using the Kruskal coordinates $U = e^{ -(t-r^*)/4GM}, V= e^{(t+ r^*)/4GM}$ such that 
\be
UV= \left(\frac{r}{2GM} -1\right) e^{-r/2GM}
\ee
the metric is defined thus
\be
ds^{2}=-\frac{32 (GM)^{3}}{r}e^{-r/2m}dUdV+r^{2}d\Omega
\ee
In these coordinates the future horizon comprises of $U=0$ surface. In the given figure of the 
collapsing star the region BDC is inside the horizon. If we take the modes as those which are 
positive frequency wrt $\partial/\partial U$ 
\cite{dewitt, unruh}. (U reverses sign across the horizon) 

We identify these modes as $p_{\omega l m}$ which are such that $\partial_U p_{\omega l m}= -i \omega p_{\omega lm}$. 

Thus straddling the horizon the $r_A$ surface has (this is almost null surface)
\be
\phi (r_A) = \sum_{\om lm} [a_{r_A} p_{\om lm} + a^{\dagger}_{r_A} p^* _{\om lm}]
\label{eqn:ra}
\ee
 According to \cite{unruh},
\be
p_{\om lm} = \frac{1}{[2 \sinh (4 \pi M \om)]^{1/2}}[ e^{2 \pi M \omega} u_{\om lm} + c.c.]
\label{eqn:past}
\ee

\begin{figure}
\begin{center}
\includegraphics[scale=0.4]{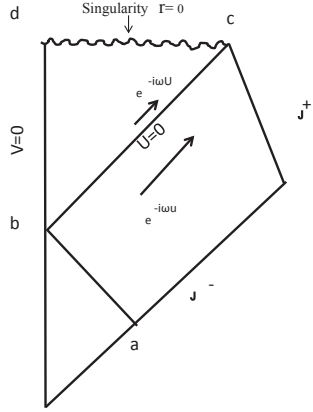}
\end{center}
\caption{Pair Creation}
\label{fig:kruskal}
\end{figure}

Using this, we find that $<U|S_{\rm quasi}|U>$ has an imaginary term which is finite at the horizon! Thus putting 
the (\ref{eqn:rb},\ref{eqn:rb1},\ref{eqn:ra},\ref{eqn:past}) in (\ref{eqn:ima}) we have in the limit 
$r_A=r_B\rightarrow 2GM$  or $\delta\rightarrow 0$
\be
W= {\rm Lim}_{\delta\rightarrow 0} 2 \ {\rm Im}\left[\int_{r_{B}}d^{3}x\sqrt{-g}g^{rr}\sum_{\omega lm}u_{\omega lm}\partial_{r}u^{*}_{\omega lm}-\int_{r_{A}}d^{3}x\sqrt{-g}g^{rr}\sum_{\omega lm}p_{\omega lm}\partial_{r}p_{\omega lm}^{*}\right]
\ee

\be
W= \int d^3 x \sin \theta \frac{1}{2 \pi }\sum_{\om lm}\frac{1}{e^{\om/T_H}-1} |Y_{lm}|^2
\ee
where $T_H= 1/8\pi M$ is the Hawking temperature. This is precisely the rate of particle emission expected and predicted in \cite{cmp}. Thus the 
flux naturally emerges as in a Bose-Einstein spectrum and is of the form as expected at the horizon! A fraction of this will emerge at 
infinity due to scattering from the exterior geometry of the black hole. This computation of the quasilocal vacuum polarisation 
amplitude is a new insight in the method of particle creation at the horizon. 
\section{Discussions}
In the previous section we showed how the expectation value of a `quasi-local' action at the horizon of a black hole shows that the scalar vacuum decays
into a flux of particles at a thermal temperature.
The interesting aspect of this is that instead of a renormalisation prescription, the difference of the flux at the inner surface and the
outersurface of the timelike cylinder emerges naturally giving the expected Hawking flux. One could thus interpret this as the total flux created
due to vaccum polarisation at the horizon. However, this derivation still does not answer the question of how gravity reacts to the escaping flux.
To study backreaction one has to quantise the matter gravity coupled system and observe the behavior of the evolution of the space-time.
A semiclassical derivation of time evolving black hole space-time exists in the framework of loop quantum gravity \cite{adg,adg1} and this requires
a non-unitary evolution. Thus we need further work to really describe the phenomena of Hawking radiation.
The above calculation can be easily extended to include Fermion propagation near the horizon. In exact analogy with the scalar action, we take the
action for the fermion field $\int d^4 x \sqrt{-g} \  \bar\psi \ e^{\mu}_a\gamma^a\left(\partial_{\mu} -\omega_{\mu}\right)\psi$
where $e^{\mu}_a$ are the tetrads and $\omega_{\mu}$ are the corresponding spin connections. Using the solutions for the Dirac modes which are positive
wrt the affine generator of the past horizon and those positive frequency wrt Schwarzschild time, one should be able to derive the boundary contributions $\int_{r_A, r_B} d^3 x ~ \sqrt{-g}~ \bar\psi\gamma^{\mu}\psi ~ n_{\mu} $ which yield the
Hawking distribution. This will be reported elsewhere \cite{mscthesis}.\\
\noindent
{\bf Acknowledgement} We would like to thank D. Page, S. Das and S. Patitsas for discussions. This work is supported by NSERC and research funds of
University of Lethbridge.

\end{document}